\documentclass[conference]{IEEEtran}
\IEEEoverridecommandlockouts
\usepackage{cite}
\usepackage{amsmath,amssymb,amsfonts}
\usepackage{algorithmic}
\usepackage{graphicx}
\usepackage{textcomp}
\usepackage{xcolor}
\def\BibTeX{{\rm B\kern-.05em{\sc i\kern-.025em b}\kern-.08em
    T\kern-.1667em\lower.7ex\hbox{E}\kern-.125emX}}
\begin{document}

\title{Learning low-frequency temporal patterns for quantitative trading}

\author{\IEEEauthorblockN{Joel Da Costa}
\IEEEauthorblockA{\textit{Department of Statistical Sciences} \\
\textit{University of Cape Town}\\
Cape Town, South Africa \\
joeldacosta@gmail.com \\
https://orcid.org/0000-0001-7821-6635}
\and
\IEEEauthorblockN{Tim Gebbie}
\IEEEauthorblockA{\textit{Department of Statistical Sciences} \\
\textit{University of Cape Town}\\
Cape Town, South Africa \\
tim.gebbie@uct.ac.za \\
https://orcid.org/0000-0002-4061-2621
}}

\maketitle

\begin{abstract}
We consider the viability of a modularised mechanistic online machine learning framework to learn signals in low-frequency financial time series data. The framework is proved on daily sampled closing time-series data from JSE equity markets. The input patterns are vectors of pre-processed sequences of daily, weekly and monthly or quarterly sampled feature changes. The data processing is split into a batch processed step where features are learnt using a stacked autoencoder via unsupervised learning, and then both batch and online supervised learning are carried out using these learnt features, with the output being a point prediction of measured time-series feature fluctuations. Weight initializations are implemented with restricted Boltzmann machine pre-training, and variance based initializations. Historical simulations are then run using an online feedforward neural network initialised with the weights from the batch training and validation step. The validity of results are considered under a rigorous assessment of backtest overfitting using both combinatorially symmetrical cross validation and probabilistic and deflated Sharpe ratios. Results are used to develop a view on the phenomenology of financial markets and the value of complex historical data-analysis for trading under the unstable adaptive dynamics that characterise financial markets. 
\end{abstract}


\section{Introduction}

\subsection{Technical Analysis}\label{intro_technicalanalysis}

Technical analysis is a financial analytical practice that makes use of past price data in order to identify market states and forecast future price movements based on past movements. The techniques typically rely on past market data (price and volume), rather than company assessments using fundamental analysis. We explore the idea that technical analysis has merit in exposing market inefficiencies when they are signified by repeated feature time-series patterns \cite{Murphy, LoHeretics}. 

Financial markets have been shown to be complex and adaptive systems where the effects of interaction between participants can be highly non-linear \cite{Arthur}, but they may also have combinations of top-down and bottom-up sources of information and interaction that mix in vast numbers of interactions mediated by numerous flocks of heterogeneous strategic agents that constitute modern financial markets \cite{WilcoxGebbie}. Complex and dynamic systems such as these may often exist at multiple `order-disorder borders' - and they will then generate certain non-random patterns and internal organisation on different averaging scales. Two key price generation processes have emerged: the low-frequency domain (the result of sequences of closing auctions generating prices), and the high-frequency intra-day domain driven by order-flow itself. Here we consider low-frequency daily sampled data that is the result of the price discovery from closing auctions.

Even at low-frequency, identifying patterns and structure is simultaneously reasonable and notoriously difficult. While it is often clear in hindsight that patterns exist, the amount of noise and non-linearity in the system can make prediction challenging. Fittingly, neural networks are a popular choices for modelling within financial markets because of their ability to perform well as universal approximators \cite{Hornik}.

Practical approaches to money management within the realities of adapting and changing market systems increasingly favour online methods, in particular \cite{Loonat} explored the application of online learning models in this space in the South African market to show that direct (but simplistic) online pattern-learning is able to identify and potentially exploit trading opportunities on the JSE through the assessment of Open High Low Close (OHLC) data. This was extended by \cite{MurphyGebbie} to more directly explore the use of online learning applied to optimizing parameters for traditional technical trading indicators as applied to maximising wealth trading zero-cost portfolio strategies.

The work presented here fits into the growing body of work which considers mechanistic and brute-force approaches of applying machine learning models to financial market data. The complexity, non-linearity, noise and stability of financial markets are highlighted through both the successes and challenges found in training these models. These difficult dynamics, and their notable difference when compared to other popular areas of ML research - which are often around Independently and Identically Distributed (IID) datasets - present fundamental problems to be explored; both in terms of prediction efficacy as well as validation. We present a framework using batch offline and online learning on JSE closing data, feature extraction and robust non-parametric validation techniques.

\subsection{Backtesting and Model Validation}\label{intro_backtesting}

Financial academic literature is currently facing a problem in terms of validation and verification of results. Trading strategy profitability has typically been proven using historical simulations, or ``backtests''. However, the recent advances in technology and algorithms available to construct these strategies have resulted in researchers being able to test increasing numbers of variations of factors. This has made it increasingly difficult to control for spurious results. The problem is so extensive that some meta-research papers suggest that ‘most published research findings are false’ \cite{Ioannidis}.

The standard way of implementing backtests is to split the data into two portions: an In Sample (IS) portion which is used to train the model, and an Out of Sample (OOS) portion which is used to test the model and validate results. If vast numbers of different model configurations are tested, then it is only a matter of time before false positives occurs with high performance both IS and OOS (i.e. overfitting) \cite{BailyPBO, McLean}. The nature of financial data makes it difficult to resolve these issues effectively. There is a low signal-to-noise ratio in a dynamic and adaptive system with only one true data sequence. Traditional hypothesis testing frameworks (e.g. Neyman-Pearson) are not sufficient in this context making more sophisticated techniques necessary.

The problem of overfitting is not novel. However, in a machine learning context, frameworks are often not suited to trading schedules with a random frequency structure. They do not account for overfitting outside of the output parameters nor take into consideration the number of trials attempted. The common `hold-out' strategy is where a certain portion of the dataset is reserved for testing true OOS performance. Numerous problems have been pointed out with this approach. The data is often used regardless, and awareness of the movements in the data may influence strategy and test design \cite{Schorfheide}. For small samples, a hold-out strategy may be too short to be conclusive \cite{Weiss}. Even for large samples, it results in the most recent data (which is arguably the most pertinent) not being used for model selection \cite{Hawkins, BailyPBO}. We present a novel application of existing sophisticated validation methods (see Section \ref{imp_assessment}) to a machine learning framework.

\section{Framework Implementation}\label{imp}

\subsection{Full Framework Process}\label{imp_fullproc}

The framework implementation brings several ideas together: i.) SAE based feature selection, ii.) Deep Learning with pre-training and weight initialization, and iii.) Online Learning and Backtest Overfitting Validation. The learning part of the framework consists of two phases: I. batch learning and II. online learning. 
\begin{enumerate}
    \item[I] {\bf Batch learning phase}: IS data is used to train Stacked Autoencoder (SAE) networks which in turn are used to perform feature reduction for Feedforward Neural Networks (FNNs) learning the price fluctuation predictions in the IS data. Both are trained using Stochastic Gradient Descent (SGD).
    \item[II] {\bf Online learning phase}: The batch trained FFN networks are used to predict price fluctuations on OOS data through Online Gradient Descent (OGD). 
\end{enumerate}
These online predictions are then sequentially used to simulate trading in a Money Management System (MMS), which in turn generates simulated returns. Finally, the MMS returns are  used as input for the Probability of Backtest Overfitting (PBO) and Deflated Sharpe Ratio (DSR) techniques in order to validate the legitimacy of the framework. 

The process has two key principles: First, implementing a generalised version of a system which could offer exploration of more complex techniques. Second, ensuring an effective modularisation of steps such that the process can be reconfigured accordingly while maintaining its integrity. In doing so, a separable system is created which brings together all key concepts. We aim to delivery the simplest implementation of a complex framework such that the effects of individual components can be properly assessed and developed. The full process flow can be seen in Figure \ref{figure-proc_diagram} in the Appendix.

\subsection{Data Processing}\label{imp_dataproc}

Datasets are transformed into log feature differences and aggregated to include the changes over rolling window periods. 
%The log feature fluctuation for measured prices $\tilde{p}$ at timepoint $i$. 
%is calculated as:
%\begin{equation}\label{eq_logdiff}
%\Delta\mathbf{p}_i = \mathrm{ln(\tilde{p}_i) - %\mathrm{ln}(\tilde{p}_{i-1})} .
%\end{equation}
The log feature fluctuations are processed for each asset's closing price at each time point $i$. Log fluctuations take compounding into account in a systematic way and are symmetric in terms of gains and losses; the log transformation also provides an ergodic time series. The rolling window summations are calculated for the past for input data {\it e.g.} (1, 5, 20) days, and in the future for predicted output - 5 days in this paper. These are calculated as summations of the log differences $\Delta\mathbf{p}_i$ such that for $d$ days at timepoint $t$:
\begin{eqnarray}\label{eq_price_agg}
\mathbf{p^{-}_{(d,t)}} = \sum_{i = t-d}^{t} \Delta\mathbf{p}_i , \mbox{ and } \mathbf{p^{+}_{(t,d)}} = \sum_{i = t+1}^{t+d} \Delta\mathbf{p}_i .
\end{eqnarray}
The aggregations are scaled using a modified Normalization, where the min and max values are determined by the Training portion of the dataset, but applied to both the Training and Prediction portions. This emulates a production implementation where future data is unknown. The log-differenced, aggregated and scaled data is then used as the input for the neural network models. Predicted outputs have the scaling and log differencing reversed in order to reconstruct the actual price point predictions for performance assessment.

\subsection{Network Weight Initialization}\label{imp_weightinit}

The problem of vanishing and exploding gradients has been one of the primary barriers to deep learning with neural networks. The approach of greedy layer wise pre-training for SAEs was suggested by \cite{Hinton1}, which allowed much deeper layered networks to be trained than previously possible \cite{Bengio1}. Once the SAE is trained, a final output layer is added and the entire network can then be fine-tuned through back-propagation without suffering such performance degradation from vanishing or exploding gradients, enabling training of both SAE and FNN networks \cite{Ranzato1, Hinton2}. 

In Section \ref{results} we see that batch training on historical data has a limited benefit, which gives primacy to weight initialization techniques for machine learning of financial time series. Initial results found that RBM pre-training for sigmoid SAE networks (as described by \cite{Hinton2}) had detrimental effects on network performance. This suggests that the IID assumptions and the different loss functions result in the financial time series data used being pathalogical for RBM pre-training, and is discussed further in \cite{JDC2020a}. It has also been shown that pre-training may largely act as a prior which may not be necessary if large enough datasets are available \cite{Bengio3, Erhan}. In the context of financial time series this prior can explain the poor performance. For these reasons we focus on variance based weight initialisations developed by \cite{Glorot} and \cite{He}. These have simpler implementations, faster computation and enables initialization for non-probabilistic activation functions, such as ReLU.

Concretely, we use a modified He initialization: ``He-Adjusted''. This initialization uses a mean of the input and output layers to scale the weight variance. For networks with constant layer sizes, the initialization is the same as He \cite{He}. For SAE networks, where layer size changes by definition, the He-Adjusted initialization results in better sized weights. For $n$, the number of nodes in a layer, we initialize using:  
\begin{equation}
\mathbf{w}_{ij} \sim U(-r, r), \mbox{ with } r = \sqrt{12/(n_i + n_j)} .
\end{equation}

\subsection{Unsupervised Learning: SAE Training}\label{imp_sae}

The benefit of the modularised system is emphasised here, as the SAE training will not suffer from limitations due to backtesting considerations: any amount of configurations can be tested for feature extraction without concern. The best chosen SAE networks (based on a minimum Mean Squared Error (MSE) score) are used to reprocess both the Training and Prediction datasets such that the input is encoded, and the output is as before. These encoded datasets can then be used for the subsequent steps in the framework. We did not implement a step to update the SAE, but results detailed in Section \ref{results} suggest that this would be an important inclusion in a production system.

\subsection{Supervised Learning: Prediction Network Training}\label{imp_ffn}

Once the predictive network is trained on IS data using SGD, the OGD process is run through the encoded Prediction dataset in order to generate the predictions for the asset prices that the model produces - thus emulating what would have occurred in a live environment.

\subsection{Money Management Strategy}

The MMS follows an arithmetic long strategy of buying any asset for which the predicted price is above the current price, and selling the stock at the prediction horizon regardless. Trading costs were included at 10\% capital costs per annum for borrowing to purchase, and 0.45\% for transaction costs as per \cite{Loonat}, without taking liquidity effects into account. The naive approach is taken purposefully so as not to bias the perspective of the system as a whole by the effects of an impactful trading strategy. It is important, in the interest of effective optimization, that the pattern prediction of the system is not tightly coupled with making it profitable. Thus, the modularity of the system is continued with a separation between the prediction signal and the MMS implementation.

%\subsection{Probability of Backtest Overfitting \& Deflated Sharpe Ratio}\label{proc_cscv}
\subsection{Validation}\label{proc_cscv}

Validation is implemented with Combinatorially Symmetric Cross-Validation (CSCV) \cite{BailyPBO} that uses the IS and OOS returns from the MMS, which in turn uses the prices from the prediction network; which is a somewhat novel application. Conceptually, the whole system comes into place here, as the results from the CSCV process are now indicative of not only backtest overfitting in the trading strategy, but also in the prediction network and without having to consider the impact of the many configuration tests for feature extraction. A modified version of ONC was run, with reduced cluster exploration. As noted in \cite{JDC2020a}, this did not appear to affect results. 

\section{Assessment Methodologies}\label{imp_assessment}

\subsection{Probability of Backtest Overfitting}\label{imp_cscv}
%\subsection{CSCV \& PBO} \label{imp_cscv}

CSCV was developed by \cite{BailyPBO} as a robust approach to assessing backtest overfitting. Their research defines backtest overfitting as having occurred when the strategy selection which maximizes IS performance systematically underperforms the median OOS performance in comparison to the remaining configurations. They use this definition to develop a framework which measures the probability of such an event occurring, where the sample space is the combined pairs of IS and OOS trading performance measures. The PBO is then established as the likelihood of a configuration underperforming the median OOS while outperforming IS. 

The CSCV methodology is generic, model-free and non-parametric, allowing it to arguably be used in any model case. By recombining the slices of available data, both the training and testing sets are of equal size (advantageous for comparing performance statistics). The symmetry of the set combinations in CSCV ensure that performance degradation is only as a result of overfitting, and not arbitrary differences in data sets. There is no requirement of a hold-out set, which removes potential credibility issues regarding whether the holdout set was treated appropriately or not. The logit distribution developed through the assessment offers a useful view on the robustness of the strategies used and the nature of the PBO score.
 
 The PBO can be estimated using the CSCV method results, which provides an estimate rate at which the best IS strategies underperform the median of OOS trials. \cite{BailyPBO} extend this to show that with models overfitting to backtest data noise, there comes a point where seeking increased IS performance is detrimental to the goal of improving OOS performance.  

\subsection{Deflated Sharpe Ratio}\label{imp_dsr}
%\subsection{DSR} \label{imp_dsr}

The Sharpe Ratio ($SR$) is based on the assumption that the returns used are the result of a single trial, as is the case for most standard performance measures. In consideration of the issues laid out in \ref{intro_backtesting}, it then becomes a misrepresentative performance measure.  \cite{BaileyFrontier} developed the Probabilistic Sharpe Ratio (PSR) which estimates the likelihood that an observed best estimated $\widehat{SR}$ exceeds a provided benchmark $SR^{*}$ (which might be expected from variance in the trials). It is worth emphasising the distinction in investment strategies between a Family Wise Error Rate (FWER), which is the  probability that one or more false positives occur, and a False Discovery Rate (FDR), which is the ratio of false positives to predicted positives. Investment strategy generations will tend to rely on the single best approach produced, and so must consider FWER. \cite{BaileySharpe} further developed the False Strategy Theorem (FST) with this in mind, allowing the assessment of whether a presented strategy is a false positive or not.

The DSR calculates the likelihood that the true $SR$ is positive under consideration of numerous trials being tested \cite{BaileySharpe}. The DSR can be estimated using the PSR methodology as $\widehat{PSR}[SR^*]$ where the benchmark Sharpe ratio, $SR^{*}$, is calculated using the False Strategy Theorem. The calculation of $SR^{*}$ requires both the variance of trial $SR$ values and the number of independent trials, which are not typically considered and where determining independence is challenging. \cite{PradoDSR} aim to resolve this with the Optimal Number Clusters (ONC) algorithm, a modified K-means methodology of clustering strategies and trial results. This clustering allows an estimation then of both the variance and number of trials, which in turn allows the DSR to be calculated. With this as a confidence level, one can accept or reject the notion that the observed $\widehat{SR}$ is positive.

\section{Experiment Process}\label{exp}

\subsection{Data \& Software}\label{exp_data}

Datasets were constructed using JSE closing price relative data for 2003-2018 \cite{JDC2020b}, with a 60:40 split on the Training:Prediction subsets. The closing price dataset consisted of 10 assets from the JSE Top 40: AGL, BIL, IMP, FSR, SBK, REM, INP, SNH, MTN, DDT (coming from a variety of sectors). More source information, data snapshots and price charts are available in \cite{JDC2020a}.

The software libraries, written in Julia, were produced for all the training, experimentation and recording of results. These are discussed extensively in \cite{JDC2020a}, and made available online \cite{JDC2020c}.

\subsection{Parameter Space Exploration}\label{proc_parameters}

The parameter space is explored using a phased grid search approach. For each stage, the relevant parameters are each specified as a set of values, and all sets are then used to generate the full combinatorial space, such that each possible combination of the specified parameters is tested.
\begin{enumerate}
	\item \textbf{Stage 1:} The data configuration (i.e. data windows, prediction point, scaling, data split points) as well as the SAE configuration (network size, learning rates, learning optimization parameters, SGD epochs) are set in Stage 1 and used to train the SAE networks.
	\item \textbf{Stage 2:} The preferred SAEs are chosen from Stage 1, and determine the data configuration used for Stage 2. These are then used to encode the datasets, which will be used for FFN training. The FFN SGD and OGD parameters are set in this stage (network size, learning rates, SGD epochs etc.), and will be combined combinatorially with the SAEs that were chosen for testing as well.
\end{enumerate}

\section{Findings}\label{results}

\subsection{Value of Historical Data and Training}\label{results_historical}

\begin{figure}
	\centering 
	\textbf{OOS P\&L by IS Training Epochs}
	\includegraphics[scale=0.29]{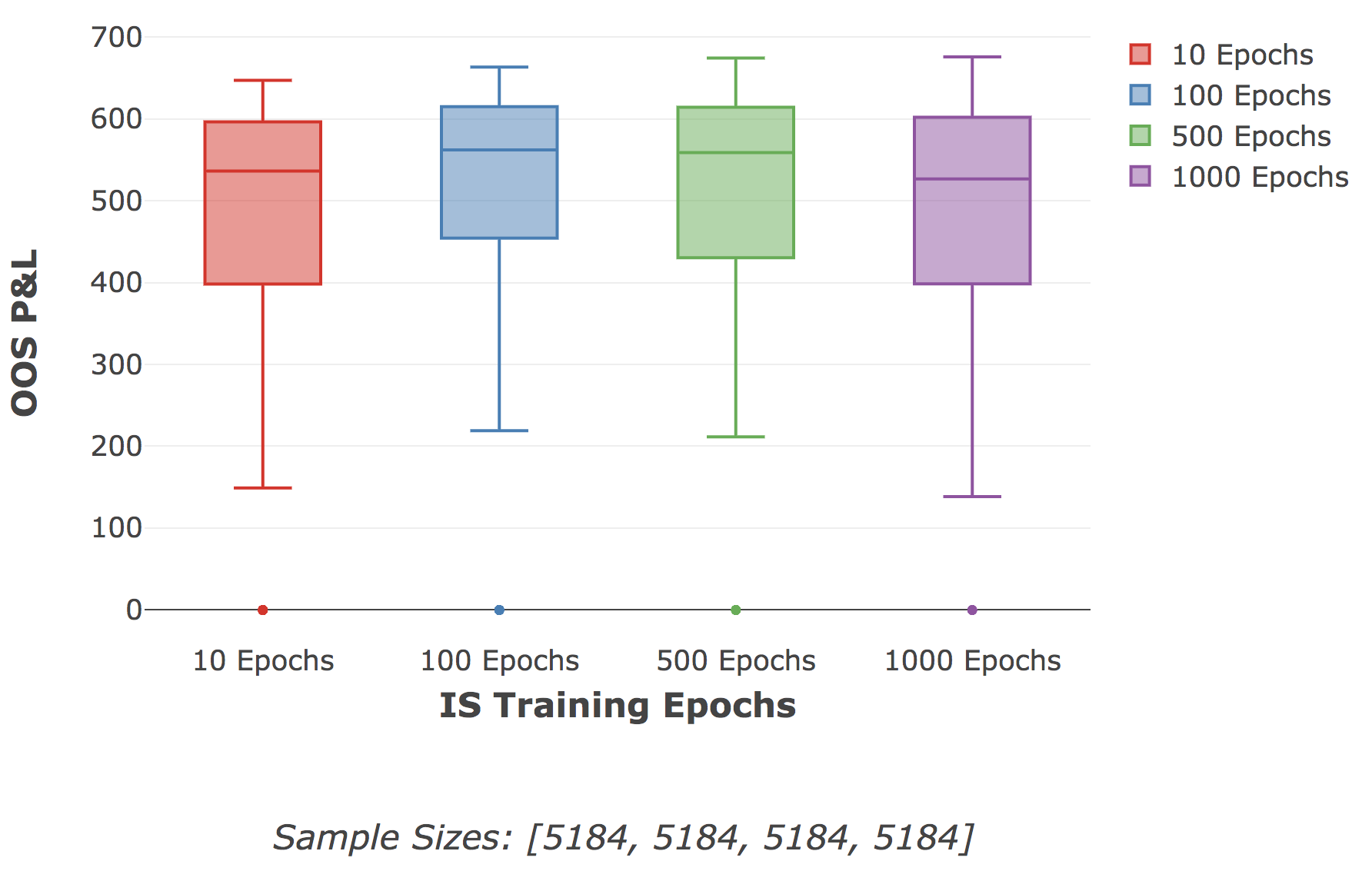}
	\caption[OOS P\&L by IS Training Epochs]
	{
		These results show OOS P\&L grouped by the number of epochs in the SGD IS training phase. Here 100 Epochs offered the best overall performance, and further training to 500 or 1000 epochs degraded performance due to the network overfitting on the IS data. The results show that the benefit of historical data is limited - having networks become better at learning return relationships from 10 years in the past did not lead to increased OOS P\&L for more current data. The small difference in the upper half of observations between 10 and 100 Epochs further emphasises this point.
	}
	\label{figure-results_pl_max_epochs}
\end{figure}

\begin{figure}
	\centering 
	\textbf{OOS P\&L by IS Training Dataset Size}
	\includegraphics[scale=0.29]{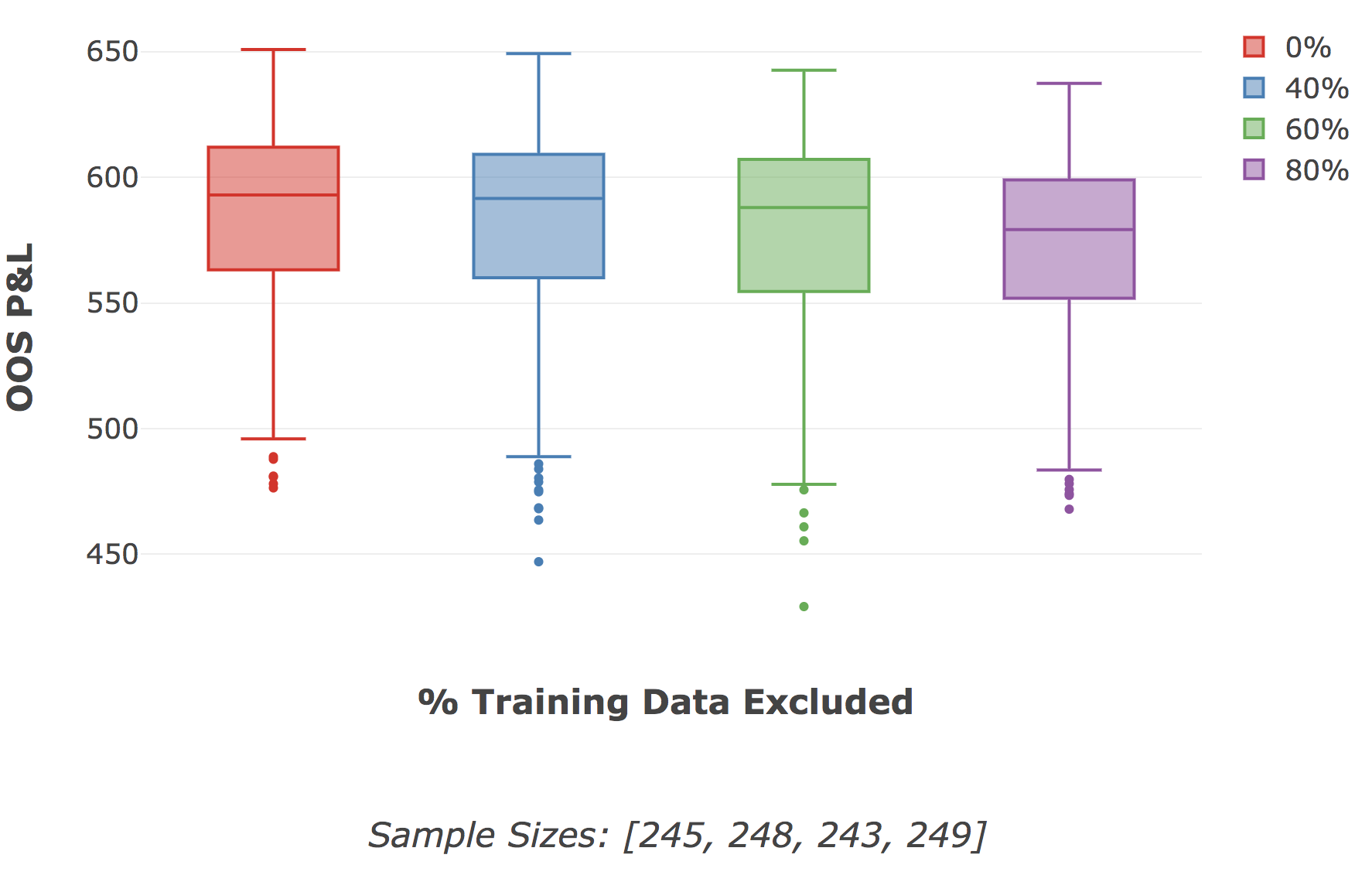}
	\caption[OOS P\&L by IS Training Dataset Size]{
		To further explore the effect of IS training on historical data, configurations were run with a percentage of the usual training data excluded, with the P\&L results grouped above. The exclusion of up to 80\% of the IS training data resulted in only a 2.2\% drop in median OOS P\&L for those networks. The training in these instances was not adjusted to increase the number of epochs according to the size of IS data, and so the configurations with more data excluded were also in essence trained less. 
	}
	\label{figure-results_it3_validationset}
\end{figure}

We expected that the IS batch training using SGD for the predictive network would improve OOS P\&L performance. Theoretically, the training on historical data might prime the network for predicting future data. However, we found that the effects of IS training had limited benefit. We ran experimental trials to test the hypothesis that the amount of historical IS data available is of limited benefit. We found the P\&L results validate this idea, as seen in Figures \ref{figure-results_pl_max_epochs} and \ref{figure-results_it3_validationset}. We saw that extensive training on past data may be akin to pre-training network weights at best, and counterproductive in overfitting to dynamics that no longer exist at worst. This highlights the complexity and dynamic nature of financial time series, where past relations and behaviours are not necessarily indicative of present state. It follows that the primary determinants of OOS P\&L are those present in the OGD (OOS) learning phase: the OGD learning rate, the data horizon aggregations, and the SAE feature selection.

This fits well with research showing that online algorithms typically perform as fast as batch algorithms during the `search' phase of parameter optimization, but that `final' phase convergence tended to fluctuate around the optima due to the noise present in single sample gradients \cite{LeCun, Bottou2}. \cite{Bottou} showed that it is actually more practical to consider the convergence towards the parameters of the optima, rather than the optima itself (as defined by the cost function) - the difference between the learning speed and optimization speed, respectively. Online learning methods are thus well suited to financial market modelling using neural networks. They allow effective and efficient incremental updates as more recent (and relevant) data becomes available. Further, the increased learning speed over optima convergence makes them a fitting choice when data is non-IID and constantly changing.

\subsection{Primary Determinants of P\&L}\label{results_primarydeterminants}

Input data was scaled to 3 different configurations to assess the effects of shorter and longer data horizons, using SAE MSE and predictive OOS P\&L to assess performance. The configurations tested (in trading day window periods) were: 1.) horizon-tuple 1 with [1, 5, 20], 2.) horizon-tuple 2. with [5, 20, 60], and 3.) horizon-tuple 3. with [10, 20, 60]. The SAE networks were only trained on IS data, and not updated afterwards. As noted below, an effective SAE feature selection in this process is an optimization that may be limited to a certain time period and may not generalise well OOS. We also found that lower variance, in the shorter horizon aggregations, resulted in easier replication; while longer horizons are more difficult (as indicated through MSE scores). The reproduction differences are discussed in \cite{JDC2020a}.

\begin{figure}
	\centering 
	\textbf{OOS P\&L By Feature Size and OGD Learning Rate}
	\includegraphics[scale=0.3]{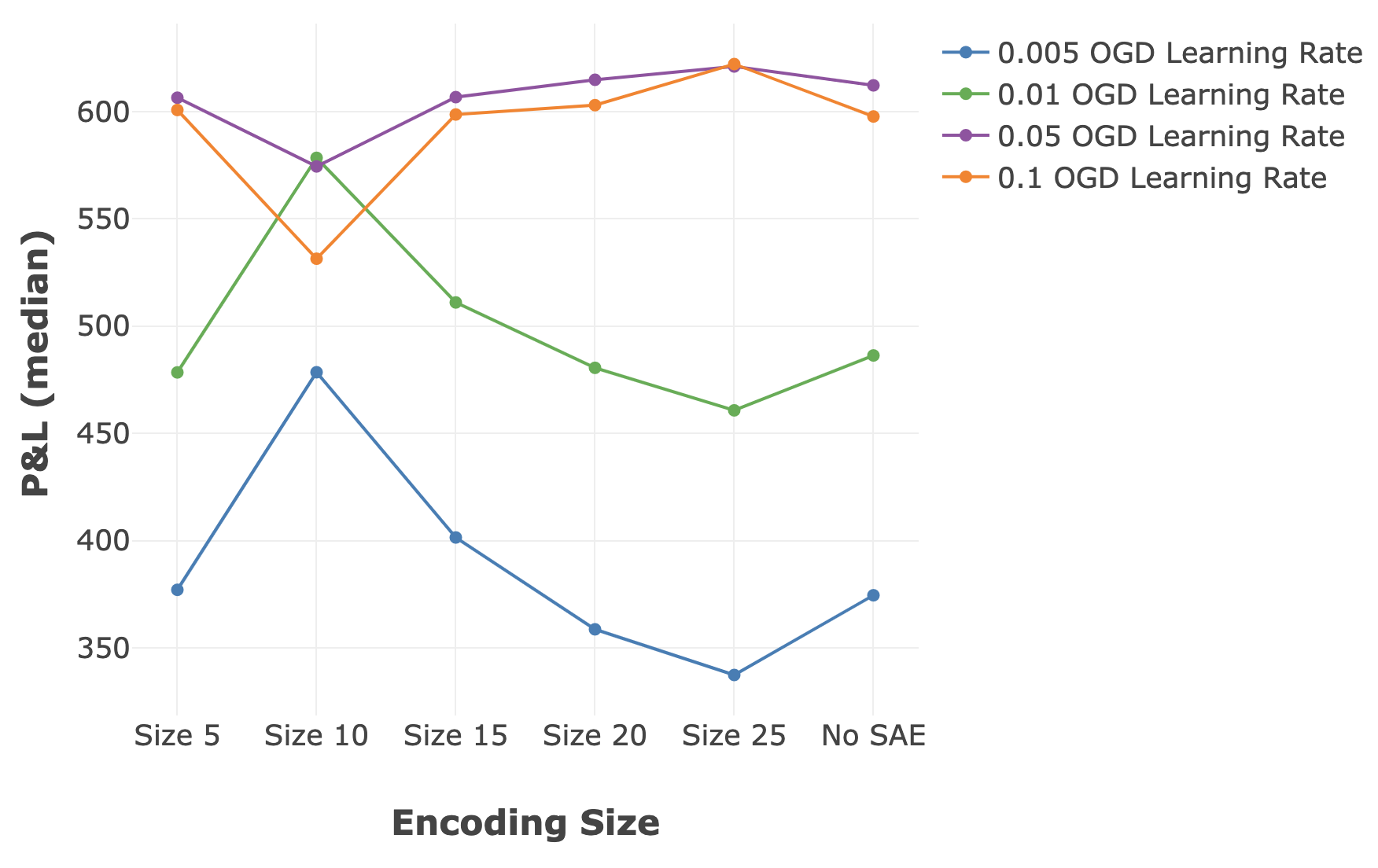} 
	\caption[OOS P\&L by Encoding Size and Learning Rate]
	{
		This figure shows that the lower learning rates (0.005, 0.01) performed best with strategies using long term trend pricing. The 10 feature encoding appeared to optimise specifically for this perspective. The optimisation caused outperformance at the lower learning rates and detrimental performance at higher learning rates (which perform best with short term fluctuation strategies). The 15 to 25 horizon encodings showed a better association to the short term strategies. Here higher encodings and higher learning rates offer the best performance. The 5 feature encoding offered consistent performance across learning rates and shows the learning of generalisable features. }
	\label{figure-encoding_pl_median}
\end{figure}

We observed strong interactions between the SAE feature sizes, FNN OGD learning rates and data horizon aggregation configurations. The performance differences seen further emphasises the unstable nature of financial systems. Generally, the FNN OGD learning rate had the largest impact on OOS performance, and demonstrates the benefits in being able to adapt quickly to new information, as seen in Figure \ref{figure-encoding_pl_median}.  As the SAE feature size decreased from 25 to 10\footnote{Input data was 10 assets with 3 horizon aggregations each, resulting in an input size of 30 at each timestep.}, SAEs learnt longer term features (as they were increasingly unable to represent short term fluctuations). For FNN networks with larger learning rates which could otherwise adapt quickly, the increased focus on long term features caused P\&L performance degradation. For FNN networks with smaller learning rates, poorly able to adapt quickly to new information either way, there was a benefit from SAE features with an increasing representation of long term trends. The relationship is emphasised dramatically in the 10-feature SAEs, to the point that lower learning rates can be more effective in generating OOS P\&L. The P\&L performance suggests that the 10-feature SAEs overfit to long term IS features, and became pathological for short term adaptation OOS. 

The most noteworthy results were the 5-feature SAEs, where performance was often on par or better than the higher feature sizes, or no SAE at all, as seen in Figure \ref{figure-encoding_pl_median}. It is possible that the small encoding layer acts as a form of regularization, forcing the SAE to learn more consistently generalisable features. The performance of the 5-feature networks, with an 83\% reduction in input data, is clear evidence of the efficacy and potential of feature selection in financial times series.

The effect of data horizon aggregations is as expected: short term horizons (i.e. [1,5,10]) outperformed in configurations with more SAE features and higher learning rates; longer term horizons (i.e. [10, 20, 60]) outperformed in low learning rate and low feature configurations. The differentiation between these groups is seen more robustly in Section \ref{results_dsr}, where data horizon aggregations are determined to be the primary clustering attribute for trade correlations. Strategies focusing on short term predictive strategies (aggregations of [1, 5, 10]) had higher variance in returns than the longer data horizon strategies, though also the highest highest P\&L and Sharpe ratios. This again shows the benefits in focusing on recent cross-sectional data in financial markets. The differentiation between the groups is discussed more in \cite{JDC2020a}.

\subsection{Money Management Strategy Results}

The benchmark is an upper bound on performance, representing MMS returns based on perfect knowledge of future prices. The benchmark full return rate is 2.4\% with trading costs, over a period of 1555 trading days. So while the strategies' proximity to the benchmark do represent a framework success, they are not necessarily representative of a feasible market solution. Ultimately, this enforces the notion that the MMS implementation is of exceeding importance in a live trading process, and predictive accuracy is only able to achieve so much.

\begin{figure}
	\centering
	\textbf{MMS OOS P\&L Distributions, with Costs Applied}
	\includegraphics[scale=0.3]{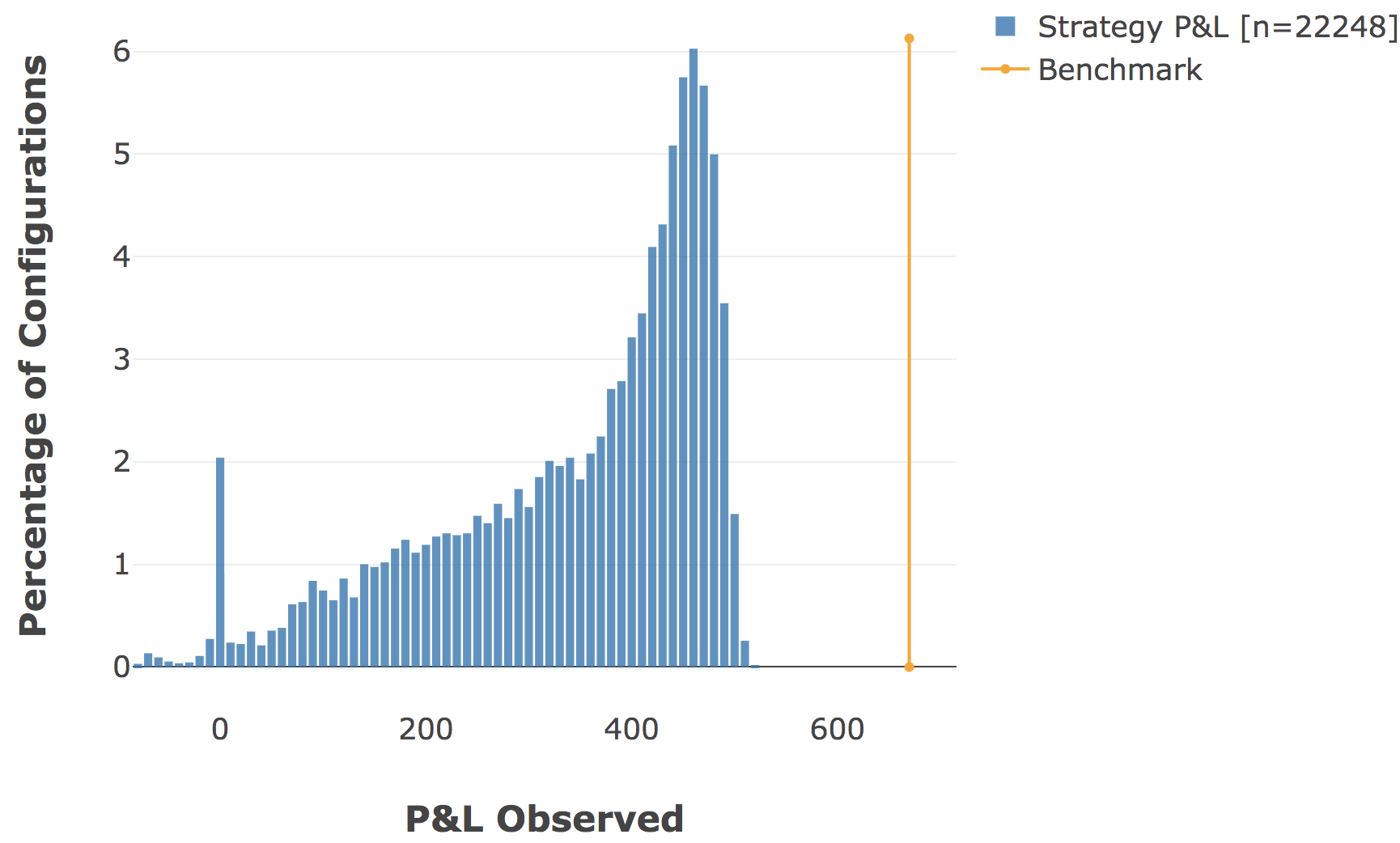} 
	\label{figure-results_pl_pdf_cost_capital}
	\caption[MMS OOS P\&L Distributions]
	{
		%Dataset: Actual10 (\ref{dataset_actual10}) ; Configuration 11 (\ref{config11}) \&  Configuration 16 (\ref{config16})
		The distributions of all OOS P\&L values, with the benchmark P\&L indicated in orange, show an encouraging view of the results. There is a significant negative skew, with a proportionally small number of strategies resulting in negative returns, even with capital and trading costs applied. There were a large proportion of strategies near the OOS upper bound, which is within 28\% of the benchmark.
	}
	\label{results_pl_pdf_cost}
\end{figure}

Figure \ref{results_pl_pdf_cost} shows the distributions of OOS P\&L with trading costs being accounted for. There were a significant number of configurations within 20\%-30\% of the benchmark. The trials with 0 P\&L are networks which suffered from either exploding or vanishing gradients, and were not able to make sufficient predictions. 

\subsection{Probability of Backtest Overfitting}

\subsubsection{Applying PBO in Mechanistic Machine Learning}

While the methodology is a model free approach to assessing overfitting, the application in a machine learning context is novel and has dynamics worth considering. The use of offline batch learning parameters, online learning parameters and adaptive network weights make the concept of model parameters less distinct. If a model performs well OOS due to effective learning, this can be due to the model's strength rather than overfitting.

It is noted that the logit metric, which the CSCV method relies on, has its basis in an ordinal ranking; indicating whether the best strategy in the IS set is higher than the median in the OOS set. This means that
poor performing configurations can artificially bolster an ordinal position past the median point and so bias PBO results. An honest, wide exploration of the parameter solution space in a mechanistic machine learning framework is likely to result in ``poor" configurations being tested (as visible in the `0' P\&L configurations in Figure \ref{results_pl_pdf_cost}). As a result, the methodology shifts the onus onto the researcher in both handling and reporting these dynamics responsibly.

Further to this, the parameter space search methodology (Section \ref{proc_parameters}) may also result in a lower likelihood of PBO due to the way of combining parameters across IS and OOS stages. By way of example, any configuration which performs well IS will have all possible OOS parameters tested in combination with it. While some of these combinations may result in poor performance, there will always be a combination of the best IS and best OOS parameter choices. This makes it unlikely that the best configurations will be past the median point for the logit calculation, resulting in a systematically low PBO regardless of how many configurations are attempted.

Lastly, the CSCV algorithm requires a parameter choice of how many windows to split the data into. While not inherently problematic, this choice can have a significant impact on results which is not visible in the reported PBO value. We discuss this further in \cite{JDC2020a}.

\subsubsection{PBO Results}\label{results_pbo_stats}

We ran the CSCV algorithm on the majority of the configurations tested, which resulted in a final PBO value of 1.7\%. A subset of networks were excluded on accounts of `null' predictions, resulting in a sample size of 21653 (out of a total of 22248 configurations). The CSCV algorithm was run with a split value of 16. There were 15 years of data, making 16 a reasonable choice as the split parameter (which needs to be even). Ideally, the splits would represent shorter periods, but the exponential increase in computational time made this impractical. The full logit distribution can be seen below in Figure \ref{figure-results_logits_all}.

We found interesting dynamics around the calculation of PBO, and the configurations contributing to the figure. The configuration process went through 2 primary phases: an extremely broad combinatorial grid search, consisting of 20736 configurations; and a second much narrower search of 1512 configurations. Assessing only the configurations from the second phase resulted in a PBO score of 6.3\%, which was significantly higher than the overall PBO score. The effect here highlights important aspects of the PBO calculation. The PBO score was much higher for the configurations which were picked more specifically after having already seen a large number of results, which is correctly indicative of increased likelihood to overfit. However, the PBO score is not monotonically increasing with N, as one would expect. This is counterintuitive and is in line with the concerns regarding the effects of increasing configuration sample size.

\begin{figure}
	\centering 
	\textbf{Logit Distribution for All Configurations}
	\includegraphics[scale=0.25]{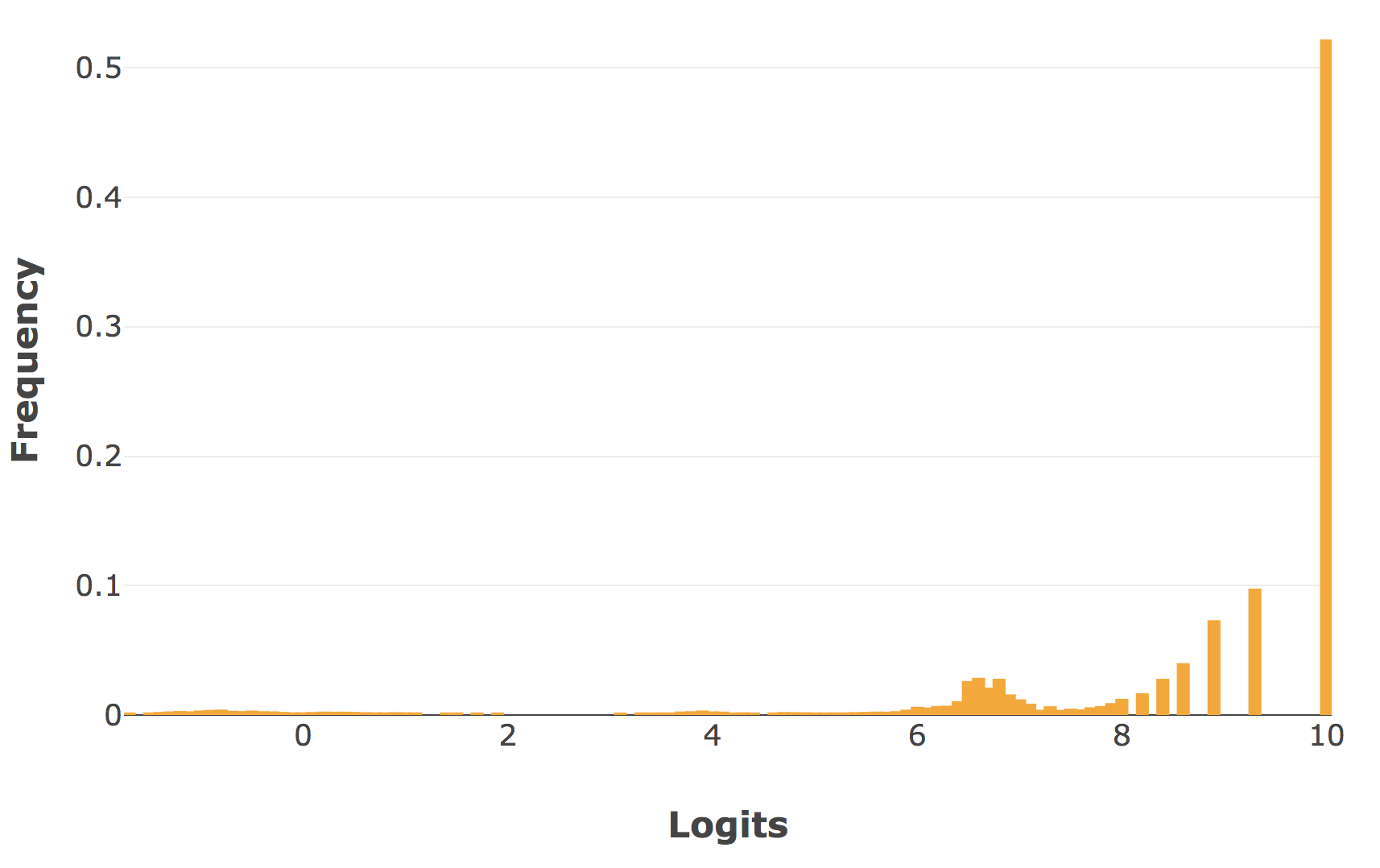} 
	\caption[Logit Distribution for All Configurations]{
		The CSCV logit distribution for the 21,653 configurations run, with a calculated PBO of 1.7\%. The strong negative skew is indicative of IS and OOS strategy returns being closely matched in rank and results in a low PBO score. This is a favourable assessment for the efficacy of the full framework presented here and shows that training was able to occur without much risk of backtest overfitting. }
	\label{figure-results_logits_all}
\end{figure}

%\subsection{ONC and DSR Results}
\subsection{Optimal Number of Clusters}\label{results_onc}

%\subsubsection{ONC Results}

The ONC algorithm produced three clusters: one consisting mostly of the negative Sharpe ratio configurations, and two containing the remaining configurations partitioned by their data horizon aggregations configurations. The distributions for all three clusters' Sharpe ratios can be see below in Figure \ref{figure-dsr_clusters}. If we consider the two primary clusters, we see that Cluster One contained all the trials with horizon aggregations of [5, 20, 60] and [10, 20, 60], and Cluster Two contained all trials with horizon aggregations of [1, 5, 10]. The nature of the experimentation process, with the combinatorial parameter space exploration (as detailed in Section \ref{proc_parameters}), is such that other parameters were mostly evenly split across the 2 clusters (e.g. OGD learning rate, network sizes, initializations and so on).

The clusters here indicate that the networks adapted to at least two different general strategies for predicting prices: one which which was more influenced by the long term fluctuations, and the second was more influenced by the short term fluctuations. The results presented in Section \ref{results_primarydeterminants} are then indicative of the networks ability to execute these overarching strategies effectively.

The best Sharpe ratio value (with trading costs applied) was 0.64 and part of Cluster Two, with the [1, 5, 10] price fluctuation horizon aggregations. The distributions seen in Figure \ref{figure-dsr_clusters} indicate that at a general level, Cluster One has more consistent performance, Cluster Two on the other hand has higher variance, with more strategies at both the lower and higher range of Sharpe ratios. The lack of further clusterings was probed manually to find that the variance from further subclustering lead to a worse cost function score. 

\begin{figure}
	\centering
	\textbf{Sharpe Ratios for ONC Clusters and Best Strategy}
	\includegraphics[scale=0.25]{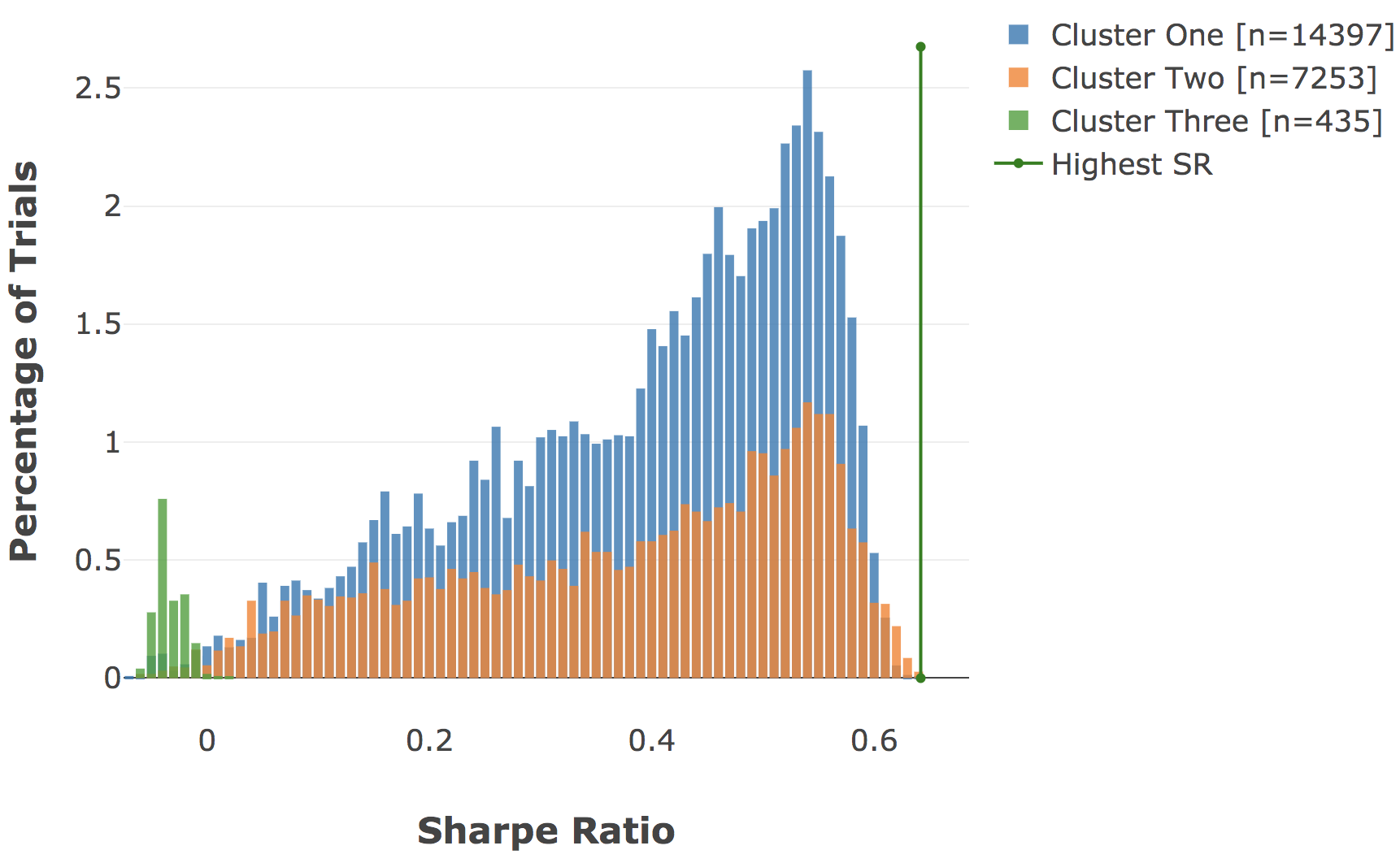} 
	\caption[Sharpe Ratios for ONC Clusters and Best Strategy]
	{
		%Dataset: Actual10 (\ref{dataset_actual10}) ; Configuration 11 (\ref{config11}) \&  Configuration 16 (\ref{config16})
		This figure shows the distributions of all Sharpe ratios, grouped by the clusters produced by the ONC algorithm, and an indication for the best Sharpe ratio (which is in Cluster Two). Cluster One has more consistent values, with a higher mean ($\mu_{c1} = 0.393$, $\mu_{c2} = 0.375$) and higher negative skewness ($\tilde{\mu}_{3,c1} = -0736.$, $\tilde{\mu}_{3,c2} = -0.543$). Cluster Two has higher variance ($\sigma_{c1} = 0.022$, $\sigma_{c2} = 0.028$) but with more strategies at both the lower and higher range of Sharpe ratios; including the highest Sharpe ratio from all trials.
	}
	\label{figure-dsr_clusters}
\end{figure}

\subsection{DSR and PSR Results}\label{results_dsr}

Using the clusters produced by the ONC Algorithm (Section \ref{results_onc}), the DSR could be determined. The aggregate cluster time series returns were calculated and annualized to allow their variance estimates to be used to calculate $SR^*$ (the maximum expected observed Sharpe ratio due to variance under the null hypothesis of $H_0:  \widehat{SR} = 0$). Using $SR^*$ as the benchmark, the PSR calculation ($\widehat{PSR}[SR^*]$) can then be used to determine if the observed $\widehat{SR}$ is a false positive or not. This gives us the DSR as a confidence for observing a positive best $SR$. The benchmark $SR^*$ calculated was 0.211245, and the best $\widehat{SR}$ observed was 0.642632, leading to a $\widehat{PSR}[SR^*]$ of 1.0, thus indicating that the trials certainly contain a strategy which has a positive SR rate. This seems a reasonable conclusion, considering the SR distributions in Figure \ref{figure-dsr_clusters}. 

\section{Conclusions} \label{sec:conclusions}

Mechanistic machine learning approaches to financial market data hold some promise for enhancing the performance of low-frequency quantitative trading. To investigate this potential we provide a novel framework that we show to be effective in both training, and in validation. The framework is configurable and based on decoupled modules and uses several well understood techniques: deep-learning neural networks for stock price fluctuation prediction, stacked autoencoders for the purpose of feature selection, both CSCV \& PBO to assess the returns from MMS, and the likelihood that backtest overfitting has taken place, and DSR in order to assess the likelihood of a positive Sharpe ratio having been observed.

While machine learning models are expected to excel in big data environments, in financial markets there is in fact a lack of data with relevant information and signal, both for training and more so for validation. We show that IS training on historical data had a limited benefit. This is not a surprising empirical insight and was confirmed by the negligible impact of increasing training time, as well as the small impact of large reductions in the training data sizes. Increased performance for IS data was not significantly linked to OOS performance. This emphasises the idea that the changing dynamics of financial markets over time need careful attention. Learning optimizations for IS training, such as regularization and learning rate schedules, were shown to have IS benefits, but little impact on OOS performance \cite{JDC2020a}. This gives weight to the importance of online learning methods for financial applications, and in turn highlights the importance of network initialization. The results showed better performance for the He-adjusted initialization and poor performance in RBM based pre-training.

The primary determinants of OOS P\&L were shown to be those which affect the online learning data and the model's ability to adapt to this. We found SAE encoding layer sizes influenced the nature of features learnt, with smaller encodings generally leading to learning of longer term features. This relationship continued from layer sizes 25 to 10, with increasing effect. The results from the 10-feature SAEs suggest the SAEs were overfit to long term IS features and pathological for OOS adaptation. The relationship changed at 5-feature SAEs, which learnt far more generalisable features. The 5-feature SAEs had very competitive performance and show that feature selection in financial time series is both possible and beneficial despite the complexity present. Predictive strategies focusing on long term changes were present in configurations with longer data horizons, less features and lower learning rates (slow adaptation). Short term strategies presented with shorter data horizons, more features and larger learning rates (quick adaptation). The data horizon was the primary separating attribute in the ONC clusters, emphasising these groupings. The short term strategies had higher variance, but also the highest returns. This again highlights both the increased value in recent information in financial markets, as well as the difficulty in using it due to the amount of noise present.  

The most challenging aspect of a mechanistic approach to learning is avoiding backtest overfitting (see Section \ref{intro_backtesting}). Probing and validating of the generalisation error was done using the PBO methodology in conjunction with DSR \cite{PradoDSR}. The results discussed in Section \ref{results} show a low likelihood of the models having overfit. The CSCV and PBO methodologies suggest that they are able to add-value to our novel implementation of machine learning models; while providing a robust assessment of results. The results were further validated using the ONC and DSR algorithms to detect positive Sharpe ratios.

The phenomenological view of financial markets based on our experimental results suggest a very limited benefit to training on long term historical financial time series. A cross sectional view of the data has far more weight in delivering OOS returns; this is noteworthy in the context of neural networks. Based on our simulation work we speculate that money management strategies can be more important determinants of OOS profitability relative to signal generation and should also be learnt.

\section*{Acknowledgment}
The authors would like to thank Sebnem Er, Patrick Chang and Turgay Celik for valuable comments on the project. 

\bibliographystyle{ieeetr}
\bibliography{IEEEabrv,JDCTG_SAEPatterns}
\newpage
\onecolumn
%\appendix
\begin{figure}
	\centering 
	\includegraphics[scale=0.55]{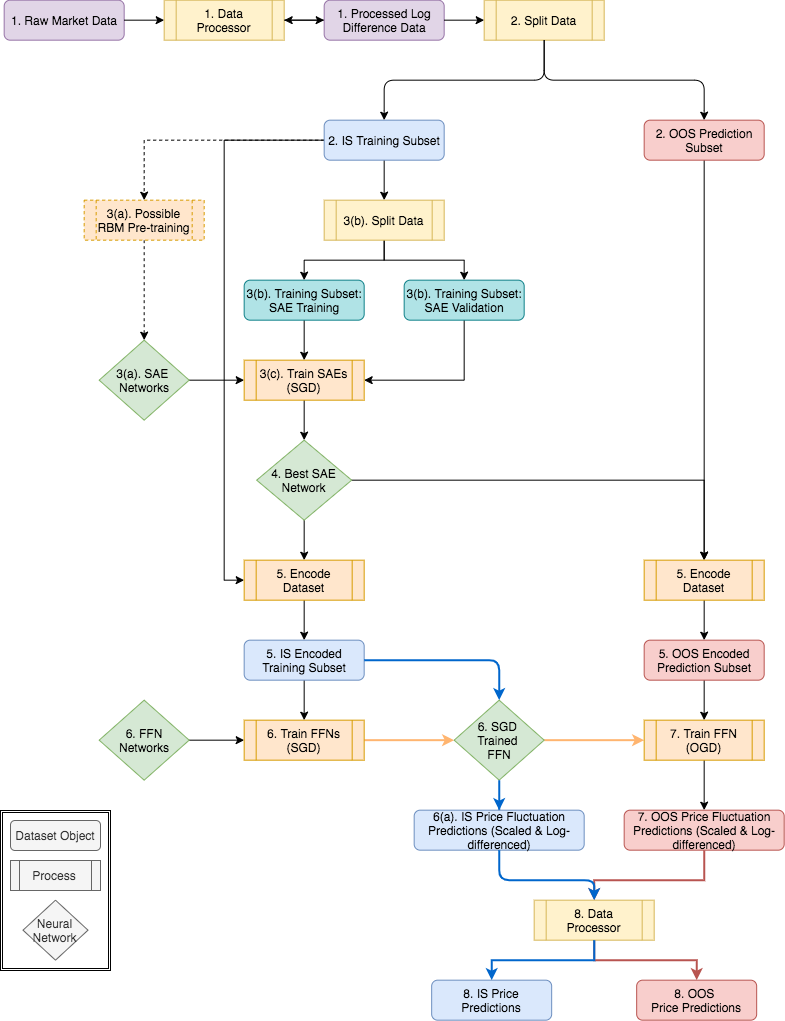}
	\caption{Overall Process Flow: The flow diagram here offers a visual representation of the framework from data processing through to training and price prediction processing, as detailed in Section \ref{imp}. The PBO \& DSR validation steps are not shown here, though use the IS and OOS Price Predictions from step 8 as inputs.}
	\label{figure-proc_diagram}
\end{figure}
\end{document}